# Molecular simulation of hierarchical structures in bent-core nematics

Stavros D. Peroukidis, Alexandros G. Vanakaras and Demetri J. Photinos

*Department of Materials Science, University of Patras, Patras 26504, GREECE*

**ABSTRACT**

The structure of nematic liquid crystals formed by bent-core mesogens (BCMs) is studied in the context of Monte Carlo simulations of a simple molecular model that captures the symmetry, shape and flexibility of achiral BCMs. The results indicate the formation of (i) clusters exhibiting local smectic order, orthogonal or tilted, with strong in-layer polar correlations and anti-ferroelectric juxtaposition of successive layers and (ii) large homochiral domains through the helical arrangement of the tilted smectic clusters, whilst the orthogonal clusters produce achiral (untwisted) nematic states.

Phase polymorphism is quite common for solids and rather rare for single-component isotropic liquids. The few known instances of liquid-liquid phase transitions are interpreted in terms of changes in the local molecular arrangement and/or molecular association [1]. In macroscopically uniform nematic fluids, on the other hand, the existence of positionally structured domains, known as cybotaxis [2], is not uncommon and is usually attributed to pretransitional fluctuations. Recently, however, it is becoming apparent that, for a class of achiral bent-core mesogens (BCMs), such local structure may persist throughout their nematic temperature range and even into the isotropic phase [3]. XRD experiments have given strong evidence that BCM nematics form cybotactic, smectic-like, clusters on a microscopic scale [4-12]. Interestingly, a case of symmetry change of the cybotaxis (from orthogonal to tilted) on varying the temperature, without a change in the macroscopic symmetry of the nematic phase, has been recently reported [4]. Furthermore, the phase stability of cybotactic nematics, including the possibility of nematic-nematic phase transitions differing in their local structure, is supported by recent theoretical studies [13-14]. A fully atomistic simulation [15] of BCMs indicates the formation of a biaxial nematic phase consisting of small polar clusters, although the small size of the sample does not permit firm conclusions on the extent and structure of the clusters.

Here we attempt to clarify the role of the basic molecular features of BCMs on the symmetry and structure of the nematic phase up to length scales involving several thousands of molecules and, thereby, to shed new light on the apparent conflicts between experimental findings such as those reported in [6] and [16]. To this end we have employed large scale Monte-Carlo (MC) computer experiments on a coarse-grained model of statistically achiral BCMs whose generic type includes the extensively studied ODBP mesogens (see Fig. 1(a)).

The model-molecules consist of four cylindrical segments (soft-core spherocylinders) connected as shown in Fig. 1(a). The two central segments (with diameter $D$ and length $L$) are joined rigidly at their end caps and form an angle $\gamma = 40°$, as suggested by molecular mechanics calculations. The two terminal segments (with diameter $D$ and length $L'$) are attached at the free ends of the central core of the molecule at an angle $\theta = 20°$. We have set $L/D = 2.3$ and $L'/D = 1.3$, in accordance with the overall molecular contour length of the actual ODBP molecule of Fig. 1(a). Despite its simplicity, this model allows us to take explicitly into account (i) the BCM architecture, (ii) the molecular flexibility, stemming primarily from the ester linkage and the end chains and (iii) the chemical differentiation between the aromatic bent core of the molecule and its flexible end-chains. A variant of the anisotropic soft-core potential, introduced in Ref. [17], is used to describe the intermolecular interaction potential between the spherocylindrical segments. Details are given in [18]. Guided by molecular mechanics calculations the intrinsic conformational statistics are conveyed by an intramolecular potential of the form $u_{Rot}(\omega) = -\varepsilon_R |\sin \omega|$, with $\omega$ denoting the dihedral angle formed by the central bent-core plane and the plane defined by the direction of the end-segment and its respective precession axis (see Fig. 1(a)). A positive $\varepsilon_R$ introduces two energy minima that correspond to molecular conformations with the end segments out of the plane of the central bent-core of the molecule. The minimum energy configurations, labeled as $C^+$ and $C^-$ in Fig. 1(a), are chiral enantiomers. Their statistical equivalence, however, renders the isolated molecule achiral.

The thermal behaviour and the molecular organisation are studied by means of standard Metropolis Monte Carlo simulations in the isobaric isothermal ensemble (N$p$T) using variable-size simulation boxes with periodic boundary conditions [19]. The pressure is held constant at the value $p = \varepsilon_0/D^3$, with $\varepsilon_0$ an energy unit that sets the scale for the reduced temperature according to $T^* = kT/\varepsilon_0$. The results presented below are for $\varepsilon_R = 4\varepsilon_0$

On cooling from the isotropic liquid (*I*) a nematic phase (*N*) is obtained, via a weak first-order transition. On further cooling, a smectic phase with fragmented-layer structure is obtained. Of particular interest is the finding that independent cooling runs from the isotropic phase do not always give a unique nematic phase. An example is illustrated by the two snapshots of Figs. 1(b) and 1(c) both obtained for the same temperature $T^* = 2.4$. The snapshot in Fig. 1(b) presents a usual uniaxial nematic state, with the molecules having their z-axes aligned on average along a unique common direction. On the other hand, the helical superstructure appearing clearly in the snapshot of Fig. 1(c), corresponds to a twisted nematic state. The temperature range of stability is the same for both nematics $(3.2 < T^* < 2.0)$. The populations of the $C^+$ and $C^-$ conformations in the uniaxial (untwisted) nematic state (to be denoted hereafter by $N^0$) are equal, whilst a clear excess of molecular conformations of a specific handedness is observed in the twisted nematic state (denoted by $N^+$ or $N^-$, according to the dominance of the $C^+$ or the $C^-$ population). The fact that the temperature of the transition from the isotropic liquid to any of the $N^0$, $N^\pm$ states is the same, to within the statistical uncertainty, indicates that, for the interaction parameterization used here, the free energies of these nematics do not differ appreciably. Furthermore, no transitions were observed between $N^0$ and either of the $N^\pm$ states, or between a $N^+$ and $N^-$ state, even after extremely long simulation runs. This is indicative of a substantial free energy barrier between the three nematic states for temperatures below $N-I$ transition temperature $T^*_{I-N}$. On the other hand, changes of the helical handedness were readily obtained in the simulations by elevating the temperature of a nematic state just above the $N-I$ phase transition and starting a new cooling sequence from there. These findings are suggestive of a relatively flat free energy landscape, with respect to the domain handedness, sufficiently close to the $T^*_{I-N}$. The strong indications that (i) the free energy minima of the different states are similar and (ii) bellow the $N-I$ transition the free energy barrier between the different states prevents thermally activated transformations between them, suggest that the structure of the phase may be strongly influenced by external stimuli (for instance: surface alignment, external electric or magnetic fields). These observations are in accord with experimental observations regarding the dependence of the domain structure of BCMs on the preparation-history of the samples [3,6].

To analyse further the differences between the nematic states obtained in the simulations, several orientational order parameters [18] and correlation functions were calculated. The biaxial order parameter [18] in $N^0$ phase is essentially zero throughout the nematic temperature range, whereas in the $N^\pm$ states the non-zero values found within thin slabs normal to helix axis is due to the broken rotational symmetry about the nematic director.

From the calculation of the usual radial pair correlation function $g(r)$ and the respective projected distributions $g_\perp(r_\perp)$, $g_{//}(r_{//})$ [19], it follows that the $N^0$ and the $N^\pm$ states exhibit purely positional correlations only over a relatively short range. These pair correlations, however, are not appropriate for detecting mixed positional and orientational correlations and therefore could not reveal clearly the presence of helical order. To confirm and analyze the helical order we have calculated the pair correlation functions

$$S_{221}^{\hat{a}\hat{a}}(r) \sim -\left\langle \sum_{i \neq j} \left[ \left( \hat{a}_i \times \hat{a}_j \right) \cdot \hat{r}_{ij} \left( \hat{a}_i \cdot \hat{a}_j \right) \right] \delta\left( r - | \vec{r}_{ij} \cdot \hat{h} | \right) \right\rangle$$

, with $\hat{a}_i$ denoting the molecular axes , $r_{ij}$ intermolecular vectors and $\hat{h}$ the helix axis (see [20] for details). In the $N^0$ state, $S_{221}^{\hat{a}\hat{a}}$ shows no structure in any direction whilst for the $N^\pm$ states $S_{221}^{\hat{z}\hat{z}}$ shows that the $z$-molecular axis is on average perpendicular to $\hat{h}$. From these calculations we estimated the pitch of the helix close to the $N-I$ transition to be $56D$ and to decrease with temperature to a value of $44D$ at $T^* = 2.2$. It should be noted, however, that a precise determination of the helical pitch is not straightforward since the observed periodicity along the helical axis is influenced by the dimensions of the simulation box through the imposed periodic boundary conditions. It is worth noting here that the onset of twisted states is closely related to the existence of a barrier between the enantiochiral conformations. Thus, systems of rigid molecules ($\theta = 0°$) or of freely rotating end-segments ($\theta \neq 0°$, $\varepsilon_R = 0$) exhibit a uniaxial nematic phase.

The correlation functions described above confirm unambiguously the nematic molecular ordering in the $N^0$ and the $N^\pm$ states as well as the chiral symmetry breaking in the $N^\pm$. To further analyze, at the microscopic level, the differences between the $N^0$ and $N^\pm$ states with respect to the local environment sensed by a single molecule, we have calculated a set of mixed positional/orientational two-dimensional pair correlation densities defined by

$$g_{l;\hat{c}}^{\hat{a},\hat{b}}(r_a,r_b) = \left\langle \sum_{i \neq j} P_l(\hat{c}_i \cdot \hat{c}_j) \delta(r_a - \vec{r}_{ij} \cdot \hat{a}_i) \delta(r_b - \vec{r}_{ij} \cdot \hat{b}_i) \Theta\left(\left(\vec{r}_{ij} \cdot (\hat{a}_i \times \hat{b}_i)\right)^2 - \sigma^2\right) \right\rangle / g_0^{\hat{a},\hat{b}}(r_a,r_b) .$$

Here $P_l$ is the Legendre polynomial of rank $l$ ($l = 1, 2, ...$), $\hat{a}_i \neq \hat{b}_i$ and $\hat{c}_i$ are axes of the $i^{th}$ molecule and $\Theta(x)$ denotes the step-function ($\Theta = 1$ for $x < 0$ change symbol and $\Theta = 0$ otherwise). The functions $g_0^{\hat{a},\hat{b}}(r_a,r_b) \sim \left\langle \delta(r_a - \vec{r}_{ij} \cdot \hat{a}_i) \delta(r_b - \vec{r}_{ij} \cdot \hat{b}_i) \Theta\left(\left(\vec{r}_{ij} \cdot (\hat{a}_i \times \hat{b}_i)\right)^2 - \sigma^2\right) \right\rangle$ give the molecular number density on the plane defined by the axes $\hat{a}, \hat{b}$ of a single molecule. In our calculations we have used $\sigma = D/2$. The complete set of calculated polar correlation functions can be found in [18]. Here we focus on the functions $g_{1;\hat{x}}^{\hat{a},\hat{b}}$ and $g_0^{\hat{a},\hat{b}}$, which provide information on the degree and the spatial extension of the polar order generated by associations of the steric molecular dipoles along the $\hat{x}$-axis of the bent-core structure. An important inference from $g_{1;\hat{x}}^{\hat{a},\hat{b}}$ and $g_0^{\hat{a},\hat{b}}$ is that polar orientational correlations, for the $N^0$ as well as the $N^\pm$ states, extend over greater distances than the orientationaly averaged positional correlations and, in addition, the range of the polar correlations is strongly anisotropic. Accordingly, each molecule in the sample can be viewed as being surrounded by a non spherical region containing neighboring molecules that have strong polar (and therefore biaxial) correlations with that molecule. The polar correlation lengths $\xi_x$, $\xi_y$ along the $x$ and $y$ molecular axes respectively were estimated from $g_{1;\hat{x}}^{\hat{x},\hat{y}}(x,y)$ as the optimal parameters for fitting the separate $x$ and $y$ dependences of this function to the functional forms $g_{1;\hat{x}}^{\hat{x},\hat{y}}(x,0) \sim e^{-|x|/\xi_x}$ and $g_{1;\hat{x}}^{\hat{x},\hat{y}}(0,y) \sim e^{-|y|/\xi_y}$. For both types of nematic states $\xi_x$ and $\xi_y$ increase with decreasing temperature, starting out with roughly equal and very small values ($\sim 1.5D$) just below the $N-I$ transition, increasing to $\xi_x \approx 8D$ and $\xi_y \approx 4D$ at temperatures deep in the nematic phase and showing clear diverging tendencies as the transition temperature to the layered phase is approached. The sign alternation of $g_{1;\hat{x}}^{\hat{x},\hat{z}}$ with $z$, together with the observation that the z-dependence of $g_0^{\hat{x},\hat{z}}(x,z)$ shows a maximum at separations close to one molecular length, indicates directly the tendency for layering with antiferroelectric (AF) order [18]. The polar correlation length along the $z$-molecular axis, $\xi_z$, is short (below one molecular length) near the $N-I$ transition. It increases up to two molecular lengths towards the low temperature end of the nematic range and exhibits a diverging tendency on approaching the transition to the smectic phase. Accordingly, for both nematic states the polarity-correlated domains share the following common features: (i) have an essentially uniaxial (about the $\hat{z}$ axis) ellipsoid shape at the high temperature end of the nematic range, (ii) for lower temperatures the size of the ellipsoids grows considerably and becomes highly biaxial, with polar correlations along the $\hat{x}$ molecular axis (direction of the "steric dipole") extending over twice the respective distance along the $y$ axis, and (iii) the size of the ellipsoid diverges in all directions, while maintaining the biaxial shape, at the low temperature end of the nematic phase. The calculated size, the anisotropy as well as the temperature dependence for these ordered domains ("cybotactic groups" in often used terminology) are in very good agreement [18] with available experimental estimates obtained from XRD studies [4,8,11].

The differentiating characteristics of the short range molecular organization in the $N^0$ and the $N^\pm$ states are primarily conveyed by the functions $g_{1;\hat{x}}^{\hat{y},\hat{z}}$ and $g_0^{\hat{y},\hat{z}}$, particularly by the location of their maxima along the $\hat{z}$ axis. The results of the calculation of these functions for a $N^0$ and a $N^-$ state at $T^* = 2.2$ are shown in Figs. 2(a) and 2(b). In both cases, these functions confirm the AF layering of the local structures. However, a tilted molecular arrangement within the layers of the cybotactic groups is found for the $N^{\pm}$ states, as opposed to the orthogonal molecular layering (SmAP$_A$ type) found for $N^0$ state. The orthogonal layering is directly evident from the symmetry $g_{1;\hat{x}}^{\hat{y},\hat{z}}(y,z) = g_{1;\hat{x}}^{\hat{y},\hat{z}}(-y,z)$ exhibited by the contours of $N^0$ (Fig. 2(b)(left)) and from the fact that the two secondary maxima (indicated by the red arrows in Fig. 2(a)(left)) are located symmetrically on the $\hat{z}$ axis at a distance of about one molecular length. In contrast, the contours of the $N^{\pm}$ states are not symmetric, i.e. $g_{1;\hat{x}}^{\hat{y},\hat{z}}(y,z) \neq g_{1;\hat{x}}^{\hat{y},\hat{z}}(-y,z)$ (Fig. 2(b)(right)) and the corresponding maxima are rotated (clockwise, in the instance of Fig. 2(a)(right)). This, together with the AF coupling of adjacent layers, suggests that the molecules within the cybotactic groups adopt an anticlinic arrangement, tilted with respect to their $y-z$ plane. Thus the internal order of the clusters corresponds to that of the SmC$_a$P$_A$ phase, which lacks a mirror plane. Representative cartoons of the local structure of the $N^0$ and $N^-$ states are given in Fig. 2(c).

Our results clearly support the existence of smectic-like clusters with in-plane polar order in the nematic phase of BCMs. The shape of these clusters is strongly anisotropic and their size grows with decreasing temperature. This cybotaxis originates from the molecular steric dipole interactions and the molecular flexibility accompanied by energetic barriers between enantiochiral molecular states. The former drives the in plane polar correlations while the latter underlies the tilt and the polar associations between adjacent layers. In our model the tilted structure of the smectic-like clusters is associated with the twisted nematic states $N^{\pm}$ whereas the untwisted state $N^0$ consists of clusters with orthogonal structure. We note here that while the in plane polar correlations are an intrinsic feature of the BCMs, the AF and anticlinic correlations between adjacent layers is a consequence of the specific rotational potential $u_{Rot}(\omega)$ used in these calculations rather than an inherently persistent property of the nematic states of BCMs. While in previous works [21] the chiral symmetry breaking of rigid or flexible achiral molecular models is dictated by microphase segregation due to the imposed strong intramolecular partioning into segments of different philicity, in our model the onset of the twisted states is due to molecular flexibility endowed with molecular homochiral conformation recognition. This recognition is amplified by the existence of a substantial energy barrier between the enantioconformers and may give rise to homochiral domains on a length scale that is orders of magnitute larger than the molecular dimensions. These findings provide a consistent basis for a unified interpretation of a number of novel properties observed in BCM nematics, including macroscopic field-induced phase biaxiality [13,16,22,23], giant flexo-electricity[24], spontaneous chiral symmetry breaking on the mesoscopic scale [5-7,20], ferroelectric switching[8] and large flow birefringence[25]. Also, the identification of a hierarchy of structures, wherein molecules can assume locally a tilted-layered arrangement to form clusters which in turn can self-organize into larger helical structures, rationalizes experimental results which appear at first sight as conflicting, such as the observation of biaxial ordering in the nematic phase of the ODBP BCMs shown in Fig. 1(a) by NMR [16] and the subsequent observation of helical superstructure formation in the same compounds [3,6]. Lastly, our results do neither demonstrate directly nor exclude the possibility of a negative bend elastic constant which, according to Dozov[26], may be an inherent feature of BCMs.

**ACKNOWLEDGMENT**

This work is funded through the EU $7^{th}$ Framework Programme under the project BIND (Biaxial Nematic Devices, #216025).

**FIGURE CAPTIONS**

FIG. 1. (a) Left: The ODBP-phOC$_{12}$H$_{25}$ bent-core molecule [16]. Middle: Coarse grained molecular model used in the simulations. Right: Edge-on views of the minimum energy chiral conformations of opposite handedness $C^+$, $C^-$, and of the achiral conformation $C^0$. Representative snapshots, calculated at $T^* = 2.4$, of (b) the untwisted nematic state $N^0$ with N=2500 molecules in the simulation box; and (c) of the twisted nematic state $N^-$ with N=3872 molecules. The primary nematic director $\hat{n}$ is depicted by the red arrow, twisting about the helix axis $\hat{h}$.

FIG. 2. Calculated correlation functions at $T^* = 2.2$ for the untwisted nematic state $N^0$ (left) and for the twisted nematic state $N^-$ (right): (a) plots of $g_0^{\hat{y},\hat{z}}(y,z)$ and (b) plots of $g_{1;\hat{x}}^{\hat{y},\hat{z}}(y,z)$. (c) Representative cartoons of the local molecular structure in the cybotactic clusters of the $N^0$ (left) and of the $N^-$ (right).

**FIGURES**

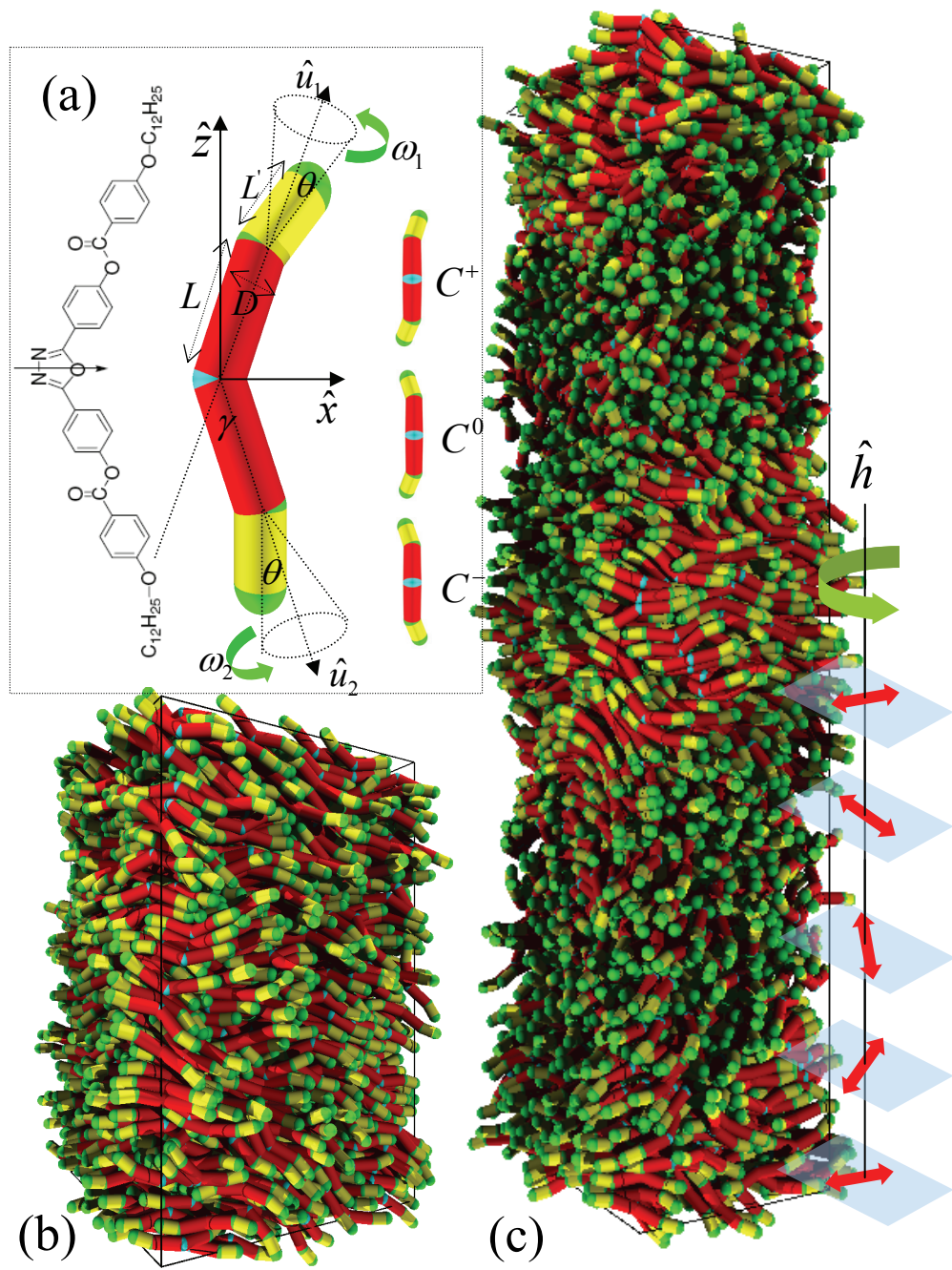

FIG. 1.

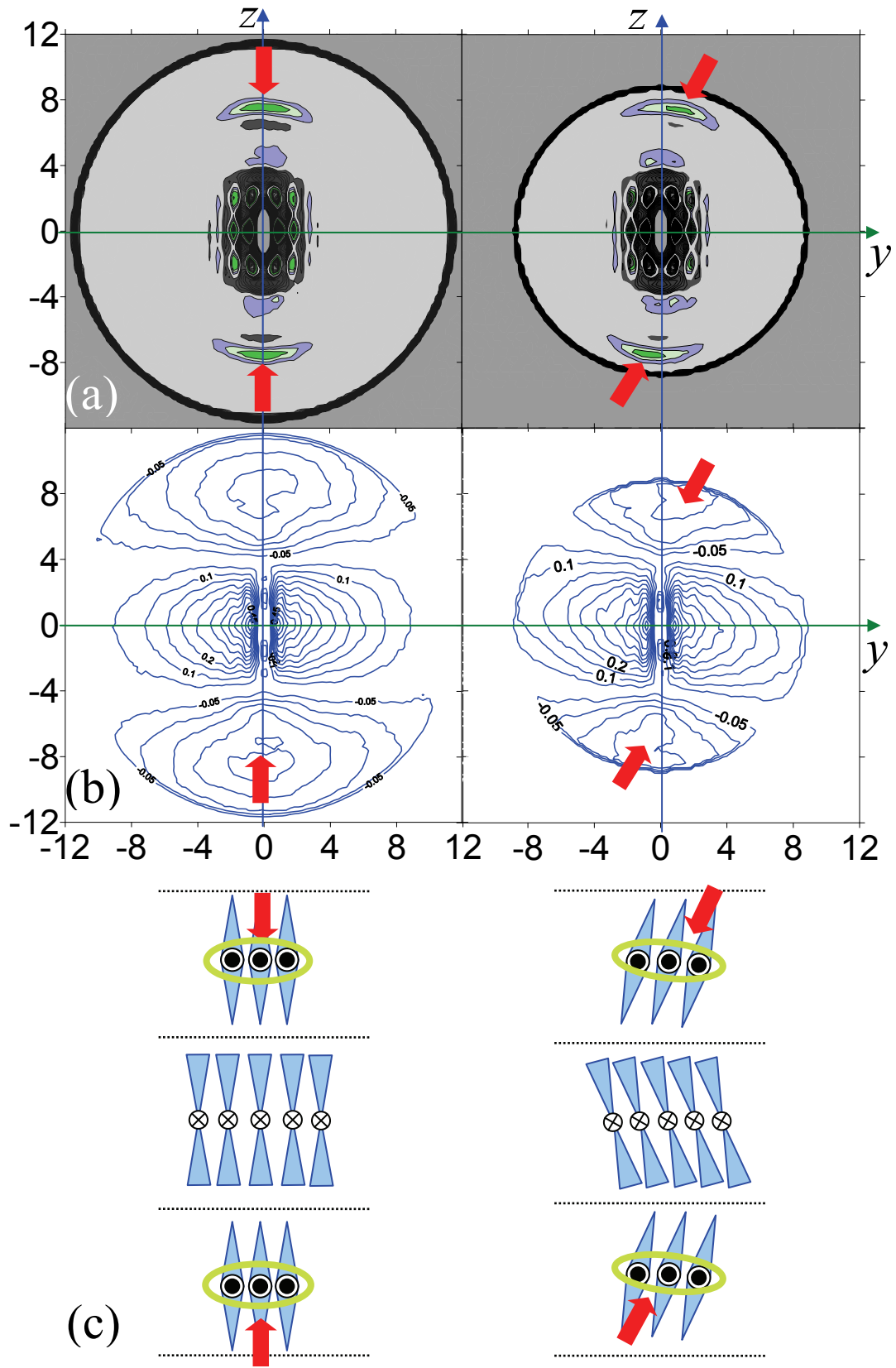

FIG. 2.